# Multishot Dual Polarity GRAPPA: Robust Nyquist Ghost Correction for multishot EPI

Yuancheng Jiang [1], Yohan Jun [2,3], Qiang Liu [4], Wen Zhong [1], Yogesh Rathi [3,4], Hua Guo [1#], and Berkin Bilgic [2,3]

[1] Center for Biomedical Imaging Research, School of Biomedical Engineering, Tsinghua University, Beijing, China

[2] Athinoula A. Martinos Center for Biomedical Imaging, Massachusetts General Hospital, Charlestown, MA, United States

[3] Harvard Medical School, Boston, MA, United States

[4] Department of Psychiatry and Radiology, Brigham and Women's Hospital, Boston, MA, United States

**# Correspondence:**

Hua Guo, PhD

Center for Biomedical Imaging Research

School of Biomedical Engineering

Tsinghua University, Beijing, China

Phone: +86-10-6279-5886

Email: huaguo@tsinghua.edu.cn

# Abstract

**Purpose:** This work aims to develop a robust Nyquist ghost correction method for multishot echo-planar imaging (EPI). The method helps correct challenging Nyquist ghosts, particularly on scanners with high-performance gradients or ultra-high fields.

**Methods:** A method for multishot EPI ghost correction, called multishot dual-polarity GRAPPA (msDPG), is developed by extending the DPG concept to multishot readouts. msDPG employs tailored DPG kernels to address high-order phase differences between two EPI readout polarities, which cannot be fully addressed using linear phase correction (LPC). Advanced regularizers can be readily employed with the proposed msDPG for physiologic inter-shot phase variation correction during reconstruction. Additionally, a calibration refinement method is proposed to improve the quality of the DPG calibration data and enhance reconstruction performance.

**Results:** Phantom and in vivo experiments on scanners with high-performance gradients and ultra-high fields demonstrated that msDPG achieved superior ghost correction performance than LPC, reducing the ghost-to-signal ratio (GSR) by over 50%. Compared to conventional DPG, msDPG provided images with lower noise amplification, particularly for acquisitions with large in-plane acceleration. Consequently, high-fidelity, submillimeter diffusion images were obtained using msDPG with regularized reconstruction.

**Conclusion:** The proposed msDPG provides a robust Nyquist ghost correction method for multishot EPI, enabling submillimeter imaging with improved fidelity.

**Keywords:** Nyquist ghost correction; multishot echo-planar imaging, diffusion imaging; dual-polarity GRAPPA

# 1 Introduction

Multishot echo-planar imaging (EPI) is a widely used technique for high-resolution diffusion-weighted imaging (DWI) and functional imaging.[1–5] Compared to single-shot EPI, it aims to improve the image signal-to-noise ratio (SNR) by shortening the echo time (TE), reduce image distortion, and increase effective resolution by mitigating relaxation-related blurring.[6,7] Recently, with the advent of advanced scanners with high-performance gradients or ultra-high fields, several studies have explored the application of multishot EPI for submillimeter diffusion or functional imaging.[8,9]

However, multishot EPI images on these advanced scanners are prone to hard-to-eliminate Nyquist ghosts.[10] Nyquist ghosts stem from differences between two EPI readout polarities, which are often caused by gradient delay and eddy currents induced by readout gradients.[11] Currently, the most widely used Nyquist ghost correction method is linear phase correction (LPC).[12] LPC uses navigator echoes without phase encoding blips to measure the differences between two polarities and models the differences as a one-dimensional (1D) linear phase term in the x-$k_y$ hybrid space. Subsequently, ghost-mitigated images can be obtained by fitting and removing this linear phase term. Although LPC works adequately on most clinical scanners, it struggles with non-linear phase differences in the readout axis on advanced scanners.[10,13–16] For example, severe eddy currents and concomitant fields on scanners with high-performance gradients can introduce high-order differences between two readout polarities,[10,15] and the increased magnetic susceptibility effects on ultra-high fields can cause severe local non-linearity.[13,14] All of these effects can cause LPC to fail, necessitating the development of improved ghost correction methods.

To provide improved Nyquist ghost correction for multishot EPI, some approaches directly estimate phase differences from the imaging data.[17,18] This can potentially outperform LPC, particularly when there are mismatches between the LPC navigator and the imaging data. However, such methods still rely on the assumption of a 1D linear phase error, limiting their ability to correct high-order differences. To address that, Chen et al. model the differences as a 2D phase error map, which achieves improved results.[19,20] Nonetheless, this 2D map is derived by fitting separate linear models along the y-axis and can be insufficient to capture high-order differences. As an alternative, it is also possible to treat data from opposing polarities as different EPI shots, employing low-rank-based regularized reconstruction frameworks like MUSSELS[4] or LORAKS[21] to implicitly estimate the 2D phase differences between the polarities.[10,22,23] While these methods demonstrate improved ghost correction performance, they generally entail a higher computational cost compared to other methods, and they double the acceleration factor per "shot" since the polarities are now split into separate virtual shots.

Recently, dual-polarity GRAPPA (DPG) has emerged as an effective ghost correction method on advanced scanners.[24,25] DPG acquires tailored calibration data and implicitly embeds 2D phase differences into multiple GRAPPA kernels. These kernels are subsequently used for concurrent ghost correction and GRAPPA reconstruction. Compared to those iterative or explicit modeling methods, DPG offers computational efficiency and has demonstrated advanced ghost correction capabilities.[23,24] However, its application is mainly confined to single-shot EPI. Although recent studies adapted DPG for multishot EPI, their methods were designed for functional imaging and relied on kernels using data across different shots, making them vulnerable to inter-shot phase variations in multishot diffusion EPI.[26,27] This limitation highlights the need for a more general DPG framework that can handle inter-shot phase variations, thus enabling its use in both functional and diffusion imaging applications.

In this work, we propose a robust Nyquist ghost correction method tailored for multishot EPI. Specifically, we extended the DPG concept to the multishot scenario, introducing multishot DPG (msDPG). msDPG decouples the ghost correction and GRAPPA reconstruction of DPG, using DPG kernels solely for ghost correction. Therefore, it functions as a pre-processing step, enabling further regularized reconstruction to enhance image fidelity and correct physiologic inter-shot phase variations. Furthermore, we developed a calibration refinement method to mitigate residual artifacts in calibration data from which the msDPG kernels are estimated, thereby improving the final reconstruction quality. The proposed method was validated using phantom and in vivo data acquired on advanced scanners via a navigator-free interleaved diffusion EPI sequence. The calibration and multishot EPI sequences were developed in Pulseq,[28] which facilitated their deployment on both the Siemens Terra.X 7T scanner and the Connectome 2.0 system with ultra-high performance gradients.[29,30]

# 2 Methods

## 2.1 msDPG

### 2.1.1 A review of standard DPG

The main concept of DPG is to simultaneously perform EPI ghost correction and image reconstruction, achieved by applying multiple sets of GRAPPA kernels across k-space. For calibration, it acquires two sets of EPI data with opposite readout polarities, whose positive readout (RO+) and negative readout (RO-) lines are separated and interleaved, forming a pair of fully sampled RO+ and RO- frames. These two frames are then combined into a ghost-free target using PLACE,[31] which involves removing the phase differences between the two frames and averaging them. Subsequently, the RO+/RO- frames and the ghost-free target are used for kernel training.

The DPG kernel training process is illustrated in Figure 1(a). As shown, multiple GRAPPA kernels (kernels A-D) are trained based on the local patterns of RO+ and RO- data. During reconstruction, these kernels synthesize both acquired and unacquired k-space lines, concurrently performing ghost correction and image reconstruction.

### *2.1.2 Formulation of DPG and msDPG for multishot EPI reconstruction*

While DPG demonstrates superior performance for single-shot EPI, its extension to multishot EPI remains challenging. First, multishot k-space trajectories contain two or more adjacent lines with the same readout polarity. Finding a suitable kernel to use these lines to synthesize a ghost-free target can be complicated. Second, although it is generally not a problem for EPI without diffusion weighting, in multishot diffusion EPI, the inter-shot phase variations present a significant challenge for GRAPPA-based reconstruction. Consequently, to employ DPG for multishot EPI in a conventional way, a common approach is first to reconstruct each shot separately using DPG. Following this, the image phase is removed on a per-shot basis, and the resulting magnitude images are then averaged, as illustrated in Figure 1(b). However, a major drawback of this strategy is the potential SNR degradation due to the g-factor penalty from applying DPG to each undersampled shot, which is exacerbated as the number of shots increases.

To overcome the aforementioned limitations, a new DPG-based method for multishot EPI ghost correction, msDPG, is proposed. Unlike conventional DPG, msDPG decouples ghost correction from GRAPPA reconstruction and utilizes DPG kernels solely to synthesize ghost-free data at the sampling locations of each shot. Therefore, it functions as a pre-processing step, similar to LPC, allowing integration with subsequent regularized reconstruction techniques (e.g., LORAKS and MUSSELS) to address inter-shot phase variations in diffusion imaging. Figure 1(c) illustrates the reconstruction process using msDPG: (i) Each shot of the multishot data is processed by msDPG to yield its corresponding undersampled ghost-free data. (ii) The individual shots are combined in k-space to form the multishot ghost-free data. (iii) Regularized reconstruction is employed to improve image fidelity and/or address inter-shot phase variations.

As an alternative approach to msDPG, a SENSE[32]-based algorithm, dual-polarity SENSE (DPS), is also developed. As illustrated and contrasted to msDPG in the Discussion, DPS yields comparable performance to msDPG. Its detailed description can be found in the Supporting Information.

## 2.2 Calibration refinement

The performance of DPG and msDPG largely depends on calibration quality, as residual artifacts in the calibration data can degrade ghost-correction performance.[33] A calibration refinement method is thereby proposed to improve calibration quality. Figure 2 presents a schematic diagram of the calibration refinement process, and its detailed steps are: (i) Divide the original calibration data into $N_{shot}$ pairs of undersampled RO+/RO- frames. (ii) Reconstruct the undersampled frames using GRAPPA, with the GRAPPA kernels trained on separately acquired gradient-echo (GRE) pre-scan data. (iii) Average the $N_{shot}$ pairs of reconstructed RO+/RO- frames, yielding a pair of mean RO+/RO- frames.

For calibration with a higher number of shots ($N_{shot} \geq 6$), a "shot combination" strategy is implemented to mitigate the g-factor penalty. Specifically, two or more shots are combined into a single shot, thus reducing the undersampling rate of each frame in steps (i) and (ii). For example, with $N_{shot} = 6$, shots (1 & 4), (2 & 5), and (3 & 6) are combined. This results in new calibration data with $N_{shot} = 3$, which are used in subsequent refinement steps.

## 2.3 Data acquisition

Phantom and in vivo data were acquired on two MRI systems: a 7T Terra.X scanner (Siemens Healthineers, Erlangen, Germany) with a Nova 32-channel head coil, and a 3T Connectome 2.0 scanner (Siemens Healthineers, Erlangen, Germany) with a custom 72-channel head coil.[34] Written informed consent was obtained from each volunteer following procedures approved by the local Institutional Review Board. The imaging sequence is a navigator-free interleaved diffusion EPI sequence with DPG and GRE calibration, implemented using Pulseq.

Two experiments were performed on Connectome 2.0:

- In Experiment 1, a basic validation was conducted. Parameters: 1×1×4 mm$^3$ resolution, 224×224 matrix, partial Fourier (pf) = 6/8, number of shots ($N_{shot}$) = 4, in-plane acceleration per shot (R) = 4, TE/TR = 30/3000 ms, echo spacing (ESP) = 0.5 ms, b = 0/1000 s/mm$^2$, number of diffusion directions ($N_{dir}$) = 3, and scan time = 2 min.
- In Experiment 2, submillimeter images were acquired using more EPI shots. Parameters: 0.76×0.76×4 mm$^3$ resolution, 288×288 matrix, pf = 6/8, $N_{shot}$ = 6, R = 6, TE/TR = 30/3000 ms, ESP = 0.68 ms, b = 0/1000 s/mm$^2$, $N_{dir}$ = 3, and scan time = 3 min.

Two more experiments were performed on Terra.X:

- Experiment 3 aimed to acquire submillimeter images. Parameters: 0.72×0.72×4 mm$^3$ resolution, 288×288 matrix, pf = 6/8, $N_{shot}$ = 4, R = 4, TE/TR = 65/3000 ms, b = 0/1000 s/mm$^2$, $N_{dir}$ = 3, and scan time = 2 min.
- Experiment 4 compared two scans with different ESPs. Common parameters were: 1×1×4 mm$^3$ resolution, 224×224 matrix, pf = 6/8, $N_{shot}$ = 3, R = 3, TR = 3000 ms, b = 0/1000 s/mm$^2$, $N_{dir}$ = 3, and scan time = 1.5 min. Specific parameters for Scan 1: TE = 60 ms, ESP = 0.83 ms. Specific parameters for Scan 2: TE = 65 ms, ESP = 1.03 ms.

## 2.4 Data processing and reconstruction

After data acquisition, geometric-decomposition coil compression (GCC)[35] was employed to reduce the number of coils. Data from Connectome 2.0 were compressed from 72 coils to 24 virtual coils, and data from Terra.X were compressed from 32 coils to 10 virtual coils. Subsequently, the proposed calibration refinement was performed on the acquired DPG calibration data.

LPC, DPG, and msDPG were used for ghost correction and image reconstruction. For DPG, each shot was first reconstructed individually, and POCS[36] was utilized for partial Fourier recovery within each shot. The magnitude images from all shots were then averaged to yield the final results. The DPG kernel size was $k_x \times k_y = 5 \times 2$.

LPC and msDPG employed SENSE-based reconstruction. They first performed their respective ghost correction step, followed by image reconstruction using the SENSE forward model with MUSSELS constraint to address potential inter-shot phase variations. The sensitivity maps were calculated from the GRE calibration using ESPIRiT.[37] The kernel size for msDPG was $k_x \times k_y = 5 \times 2$. For b = 0 s/mm$^2$ data, GRAPPA-based reconstruction was also implemented for comparison. Specifically, the data first underwent ghost correction via either LPC or msDPG. Subsequently, GRAPPA was applied to resolve the in-plane acceleration, followed by partial-Fourier recovery using POCS.

The reconstruction process was implemented in MATLAB (MathWorks, Natick, MA, USA) on a Linux workstation (CPU: 2.7 GHz, 40 cores; RAM: 512GB). After reconstruction, ghost-to-signal ratios (GSR)[38] were calculated within manually defined regions-of-interest (ROIs) to evaluate the ghost correction performance of each method.

# 3 Results

Figure S2 shows the effect of the proposed calibration refinement on the acquired calibration data. In both Connectome 2.0 and Terra.X cases, the calibration images before refinement all contain residual background artifacts, while those after refinement are mainly artifact-free. Furthermore, the noise levels in images after refinement were visually similar to those before refinement, even for the challenging $N_{shot}$ = 6 case, demonstrating the effectiveness of the proposed "shot combination" strategy.

The image reconstruction results from the 4-shot and 6-shot experiments on Connectome 2.0 are shown in Figures 3 and S3, respectively. LPC struggles in both cases and produces images with noticeable ghosts, likely due to high-order phase differences caused by large eddy currents and concomitant fields on this scanner. In contrast, msDPG shows superior performance and effectively mitigates the ghosts seen in LPC images, as evidenced by lower average GSR values in the phantom results (2.6% v.s. 5.6%). While standard DPG also helps reduce the ghosts, it suffers from g-factor penalties and results in higher noise levels than msDPG, especially in the 6-shot case.

Figure 4 demonstrates the effectiveness of the proposed msDPG on the 7T scanner. Data were acquired using 4-shot EPI on Terra.X. LPC images contain residual ghosts in both b = 0 s/mm$^2$ and DWI cases (GSR = 7.3%), potentially due to non-linear phase differences arising from increased magnetic susceptibility effects at ultra-high fields. In contrast, msDPG effectively suppresses the ghosts, yielding a lower GSR of 5.4%. Although DPG also provides ghost-free results, its images show higher noise levels than those from msDPG, due to its inability to jointly reconstruct the four shots.

Figure 5 shows DWI images acquired on Terra.X using a 3-shot EPI with ESP = 0.83 ms. The LPC results exhibit severe ghosts, probably because the ESP used (0.83 ms) falls within the "forbidden echo-spacing" range, which can potentially cause gradient mechanical resonance and exacerbate ghosts. To investigate this, another scan was conducted with ESP = 1.03 ms, and gradient resonance spectra were calculated for both scans. The results are shown in Figure S4. The LPC images from Scan 2 (ESP = 1.03 ms) contain fewer ghosts than those from Scan 1 (ESP = 0.83 ms). Furthermore, the main peak of the gradient spectrum for Scan 1 (ESP = 0.83 ms) is within the "forbidden frequency" bandwidth, while the peak for Scan 2 is outside this range. These results suggest that one probable reason for the severe ghosts in the LPC images of Scan 1 is mechanical resonance, which can be avoided by prolonging the ESP. Nonetheless, both DPG and msDPG robustly correct the Nyquist ghosts for both scans, with msDPG providing images with lower noise levels.

As a pre-processing method, msDPG is also compatible with GRAPPA-based reconstruction for acquisitions without diffusion weighting. Figure S5 shows in vivo b = 0 s/mm$^2$ images acquired on Terra.X, reconstructed using GRAPPA or SENSE-based methods.

msDPG+GRAPPA offers results comparable to those of DPG and can largely eliminate the artifacts in the LPC+GRAPPA results. Notably, SENSE results generally have fewer artifacts than those of GRAPPA, likely because the MUSSELS constraint used in the SENSE reconstruction helps address minor inter-shot variations in the multishot data.

## 4 Discussion

This study introduced msDPG, a robust Nyquist ghost correction method for high-resolution multishot EPI. A complementary calibration refinement method was proposed to improve the quality of the DPG calibration data. Experimental results on scanners with high-performance gradients or ultra-high fields demonstrated that the msDPG outperformed LPC and achieved lower noise amplification than conventional DPG. High-fidelity, submillimeter diffusion images were successfully reconstructed using the proposed framework.

The proposed msDPG adopts the same calibration acquisition and kernel training process as conventional DPG. It therefore inherits the primary advantage of DPG, implicitly capturing 2D high-order phase differences for advanced ghost correction. LPC, however, uses a simple 1D linear model to represent the differences. On scanners where high-order effects are present (e.g., strong eddy currents and concomitant fields on Connectome 2.0, and severe $B_0$ inhomogeneity on Terra.X), LPC could not model the phase differences completely and yields images with residual ghosts. In contrast, msDPG produces images largely free of artifacts. Even in extreme cases where gradient resonance causes severe ghosts in LPC images, msDPG can still robustly eliminate them (Figure 5). However, it should be noted that residual ghosts in LPC images were less apparent in the DWI images compared to the b = 0 s/mm$^2$ images. A likely explanation is that the ghost signal already has less power than the signal component at b = 0 s/mm$^2$, and the presence of diffusion gradients further attenuates both. Consequently, the attenuated ghost signal may become comparable to or fall below the noise levels in DWI scans, rendering the artifacts less conspicuous than those in the b = 0 s/mm$^2$ images.

The key difference between msDPG and the conventional DPG is the decoupling of ghost correction and GRAPPA reconstruction. While the general principle of separating these steps has also been explored by others,[27,39] this work presents a practical implementation and validation of this approach for multishot EPI. This design allows msDPG to function as a pre-processing step, enabling joint reconstruction for all shots. In contrast, conventional DPG reconstructs each shot separately, suffers from increased g-factor penalties, and yields images with higher noise levels, especially in diffusion images acquired with a large number of shots (Figure S3). Furthermore, this "plug-and-play" nature renders msDPG well-compatible with advanced regularized reconstruction algorithms. In this study, MUSSELS was used for reconstruction. Deep-learning-based methods can also be integrated to potentially further improve reconstruction quality and speed.[40]

Previously, Hoge et al.[26] and Zhang et al.[27] designed tailored DPG kernels that use data across different shots for multishot EPI reconstruction. However, this cross-shot approach is incompatible with navigator-free multishot diffusion EPI due to inter-shot phase variations, requiring a combination with navigator-based GRAPPA reconstruction methods[41–43] for phase correction. Consequently, their demonstrations were confined to applications without diffusion weighting. In contrast, the proposed msDPG employs shot-by-shot ghost correction, making it robust to the inter-shot phase and directly applicable to both diffusion and functional imaging.

Like conventional DPG, the performance of msDPG also depends on the calibration quality. Since the calibration data are usually acquired using multishot EPI, they are prone to artifacts arising from potential inter-shot variations.[44] Hence, we developed a calibration refinement method to mitigate the residual artifacts in calibration data, which achieved adequate performance in our experiments (Figure S2). Notably, Hoge et al. also proposed a calibration improvement algorithm for DPG.[24] However, their method mainly focuses on optimally combining the original RO+ and RO- data into a single, clean target, while our method focuses on providing a pair of artifact-free RO+ and RO- frames.

Our alternative method, DPS, was prototyped before msDPG. Figure S6 presents a comparison of their reconstruction results. As shown, the overall image quality and ghost levels in the DPS results are qualitatively comparable to those of msDPG, and their GSR values of phantom reconstruction are also similar (2.6% for msDPG and 2.9% for DPS). However, the later-developed msDPG generally offers greater flexibility: it is a pre-processing procedure and is thereby compatible with GRAPPA-based reconstruction (Figure S5), while DPS is intrinsically a SENSE-based method. Furthermore, msDPG has a faster reconstruction speed than DPS. On our workstation, the per-slice reconstruction times for the 4-shot Terra.X $b = 0$ s/mm$^2$ data were 32.4 seconds with LPC+MUSSELS, 42.6 seconds with msDPG+MUSSELS, and 79.9 seconds with DPS+MUSSELS.

This study has several limitations. First, the application and validation of msDPG were restricted to 2D multishot EPI. Further work is needed to adapt and evaluate msDPG for sequences with advanced encoding such as SMS-EPI[43,45] or 3D EPI[46–48], where conventional DPG has already been applied.[49,50] Second, while our validation focused on spin-echo-based diffusion imaging, the proposed msDPG framework is fundamentally compatible with gradient-echo (GRE) EPI. Future work includes extending msDPG to GRE-EPI-based BOLD fMRI.[51,52]

# 5 Conclusion

This work proposes msDPG, a robust Nyquist ghost correction method for multishot diffusion EPI on high-performance scanners. msDPG uses DPG kernels to achieve high-quality ghost correction, allowing advanced regularized reconstruction for inter-shot phase variation correction. Consequently, msDPG yields improved ghost correction and reconstruction performance compared to other state-of-the-art methods, enabling high-resolution imaging with improved fidelity.

# 6 Acknowledgment

This work was supported by research grants NIH R01 EB028797, P41 EB030006, U01 EB026996, R01 EB032378, UG3 EB034875, R01 EB034757, R21 AG082377, S10 OD036263, and NVidia Corporation for computing support.

# Figures

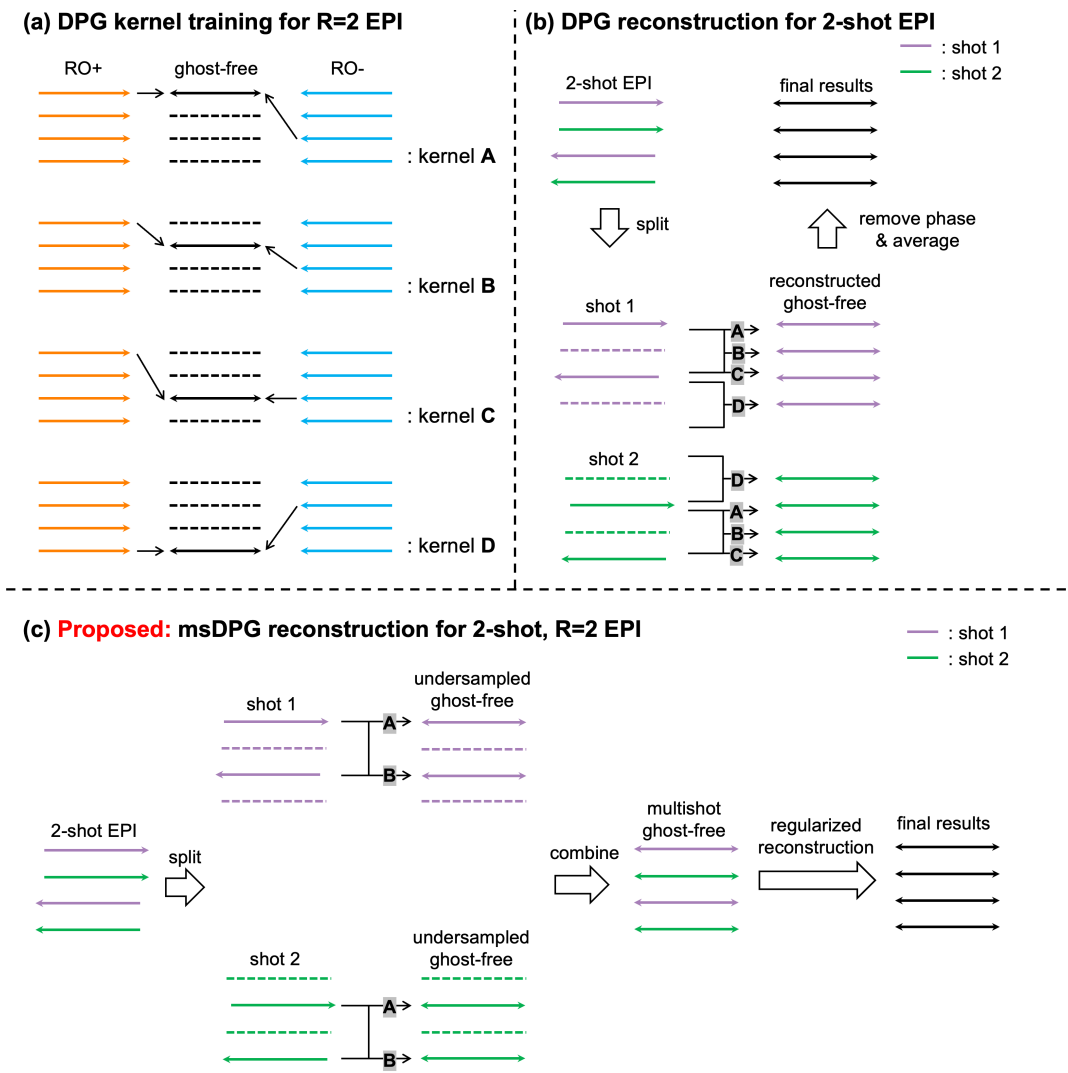

**Figure 1** Multishot EPI reconstruction using conventional dual-polarity GRAPPA (DPG) and the proposed multishot DPG (msDPG). Here, the EPI shot number is 2, and each shot is 2× undersampled. In each k-space frame, solid and dashed lines represent acquired and unacquired phase encoding lines, respectively. The arrow direction in the solid lines indicates the readout direction: rightward arrows represent positive readout (RO+), and leftward arrows indicate negative readout (RO-). Lines with double-headed arrows represent ghost-free data. (a) DPG kernel training. Multiple kernels are trained to use RO+ and RO- data to synthesize the ghost-free data. (b) Conventional DPG reconstruction for multishot EPI. Each shot is reconstructed by DPG separately. The magnitude results are then averaged. (c) msDPG reconstruction for multishot EPI reconstruction. msDPG applies DPG kernels solely to synthesize k-space data at the sampling locations, enabling subsequent regularized reconstruction.

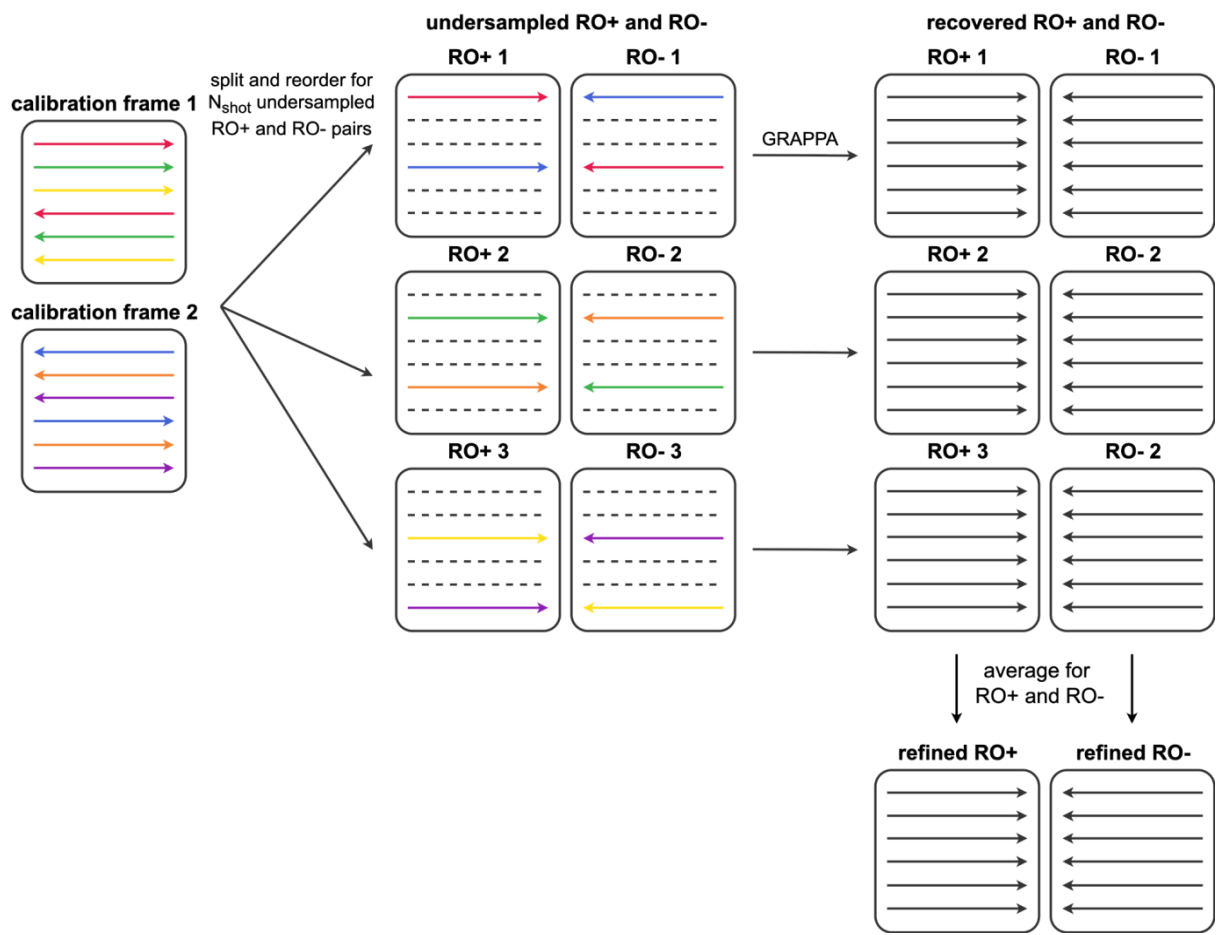

**Figure 2** Calibration refinement workflow. Without loss of generality, a 3-shot case is illustrated, with lines of different colors representing different shots. In each k-space frame, solid and dashed lines represent acquired and unacquired phase encoding lines, respectively. Arrow direction in the solid lines indicates the readout direction: rightward arrows represent positive readout (RO+), and leftward arrows indicate negative readout (RO-). The calibration data are first divided into $N_{shot}$ pairs of undersampled RO+ and RO- frames. GRAPPA reconstruction is then applied to recover these data. After that, a pair of refined RO+ and RO- frames is obtained by averaging all the $N_{shot}$ pairs.

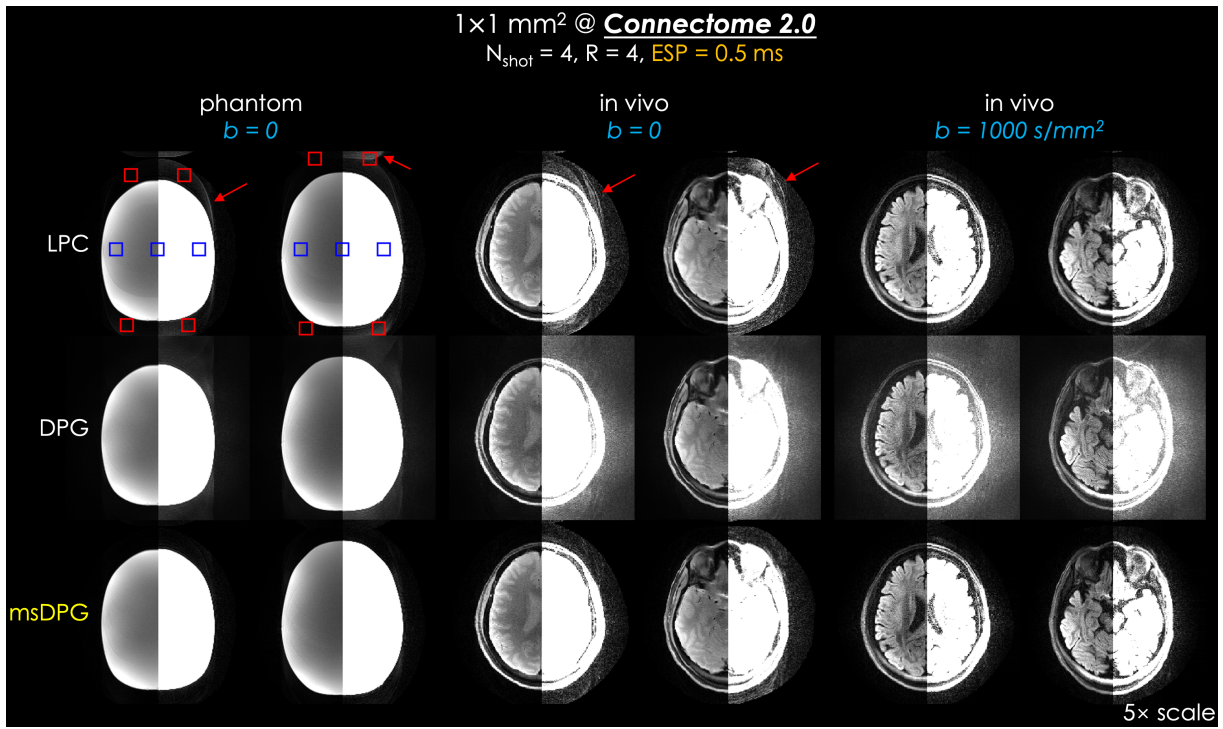

**Figure 3** Phantom and in vivo images acquired from a Connectome 2.0 scanner using a 4-shot EPI sequence. Phantom (b = 0), in vivo (b = 0), and in vivo (b = 1000 s/mm$^2$) results are respectively displayed in the left, middle, and right panels. Each panel contains two representative slices. The right half of each image is intensified fivefold to emphasize ghosts. The blue and red rectangles in the phantom images represent ROIs for signal and ghost, respectively. In phantom and in vivo b = 0 s/mm$^2$ cases, LPC reconstruction exhibits visible ghosts (arrows), whereas DPG and msDPG yield improved results. Due to the g-factor penalty, the SNR of DPG images is lower than that of msDPG, particularly in the in vivo results.

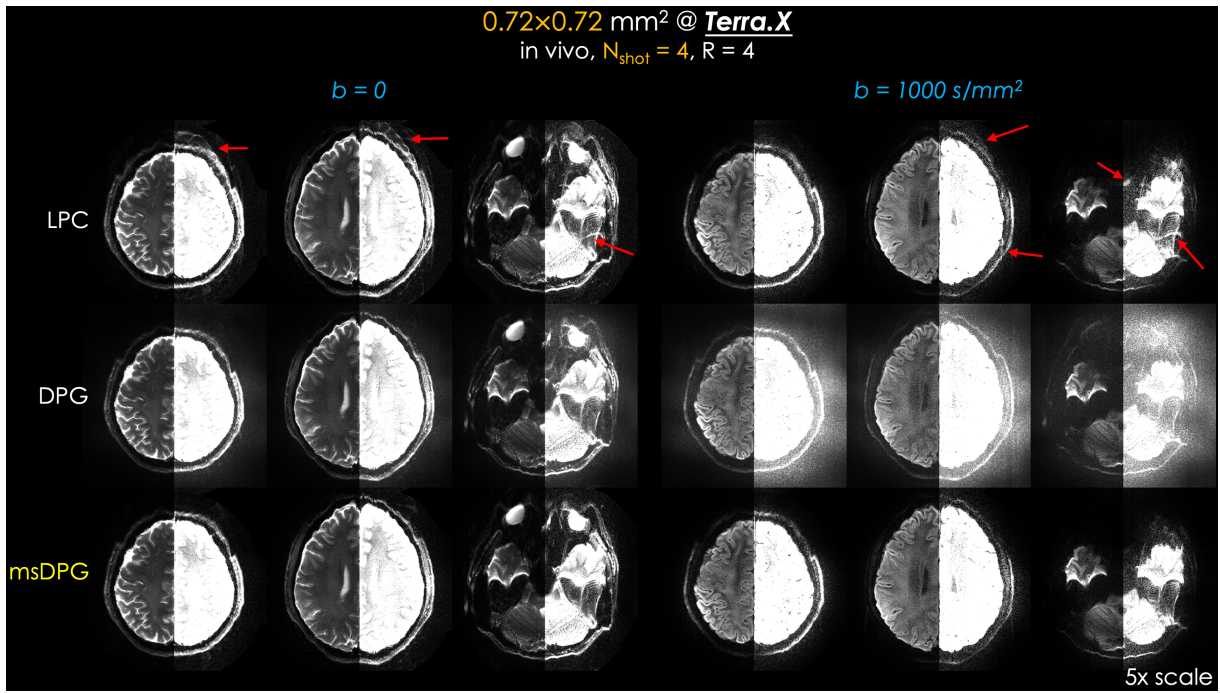

**Figure 4** Submillimeter in vivo multishot EPI images. Data were acquired from the Terra.X scanner using a 4-shot EPI sequence. The left panel shows the b = 0 s/mm$^2$ results, and the right panel shows the b = 1000 s/mm$^2$ results, with three representative slices shown in each panel. The right half of each image is intensified fivefold to emphasize ghosts. In both the b = 0 s/mm$^2$ and b = 1000 s/mm$^2$ cases, LPC images exhibit noticeable ghosts (arrows), while the images from DPG and msDPG are ghost-free. DPG produces images with lower SNR as it has no joint reconstruction capacity, especially in the b = 1000 s/mm$^2$ case.

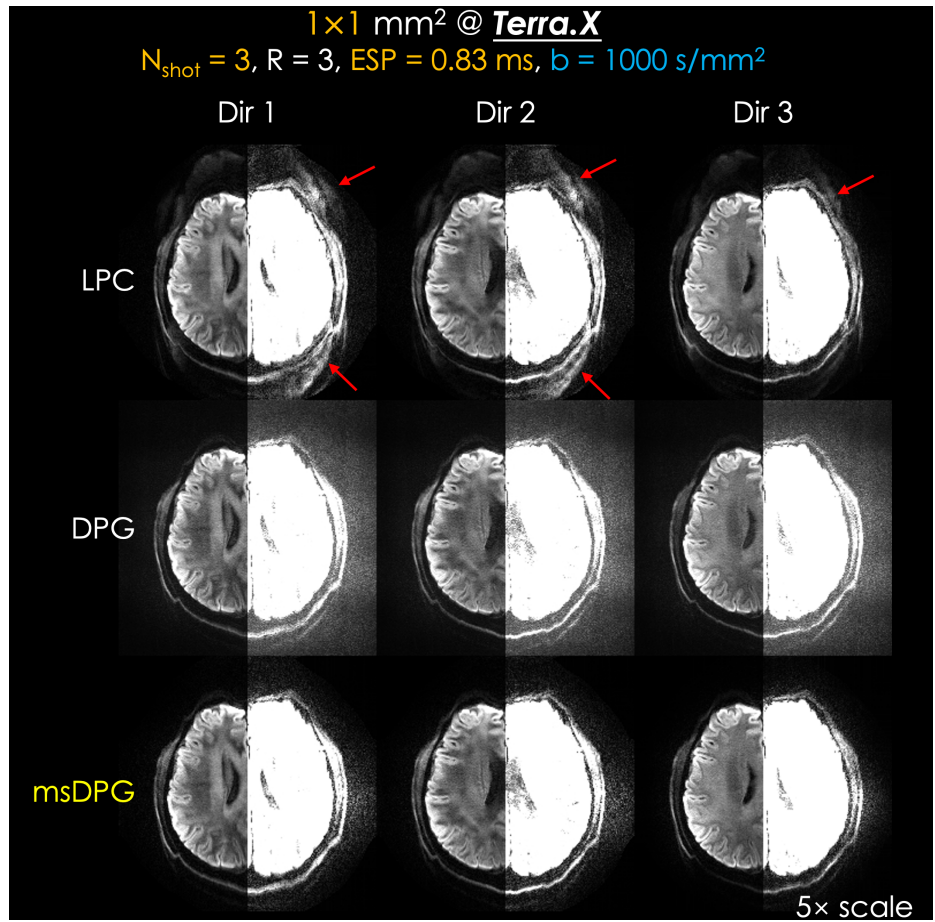

**Figure 5** In vivo diffusion images acquired on a Terra.X scanner using a 3-shot EPI sequence. The echo spacing used in this case is close to the "forbidden echo spacings," which can cause gradient mechanical resonances and exacerbate ghosts. One representative slice is displayed, with images from three different diffusion directions shown from top to bottom. LPC results contain severe ghosts across different diffusion directions (arrows). DPG and msDPG effectively mitigate the ghosts, while msDPG provides images with lower noise levels than DPG. The ghost-to-noise (GSR) values for LPC and msDPG results are 12.4% and 3.3%.

# References

1. Butts K, de Crespigny A, Pauly JM, Moseley M. Diffusion-weighted interleaved echo-planar imaging with a pair of orthogonal navigator echoes. *Magn Reson Med*. 1996;35(5):763-770.

2. Kim SG, Hu X, Adriany G, Uğurbil K. Fast interleaved echo-planar imaging with navigator: high resolution anatomic and functional images at 4 Tesla. *Magn Reson Med*. 1996;35(6):895-902.

3. Chen NK, Guidon A, Chang HC, Song AW. A robust multi-shot scan strategy for high-resolution diffusion weighted MRI enabled by multiplexed sensitivity-encoding (MUSE). *Neuroimage*. 2013;72:41-47.

4. Mani M, Jacob M, Kelley D, Magnotta V. Multi-shot sensitivity-encoded diffusion data recovery using structured low-rank matrix completion (MUSSELS). *Magn Reson Med*. 2017;78(2):494-507.

5. Niranjan A, Christie IN, Solomon SG, Wells JA, Lythgoe MF. fMRI mapping of the visual system in the mouse brain with interleaved snapshot GE-EPI. *Neuroimage*. 2016;139:337-345.

6. Wu W, Miller KL. Image formation in diffusion MRI: A review of recent technical developments. *J Magn Reson Imaging*. 2017;46(3):646-662.

7. Butts K, Riederer SJ, Ehman RL, Thompson RM, Jack CR. Interleaved echo planar imaging on a standard MRI system. *Magn Reson Med*. 1994;31(1):67-72.

8. Liao C, Bilgic B, Tian Q, et al. Distortion-free, high-isotropic-resolution diffusion MRI with gSlider BUDA-EPI and multicoil dynamic B0 shimming. *Magn Reson Med*. 2021;86(2):791-803.

9. Berman AJL, Grissom WA, Witzel T, et al. Ultra-high spatial resolution BOLD fMRI in humans using combined segmented-accelerated VFA-FLEET with a recursive RF pulse design. *Magn Reson Med*. 2021;85(1):120-139.

10. Ramos-Llordén G, Lobos RA, Kim TH, et al. High-fidelity, high-spatial-resolution diffusion magnetic resonance imaging of ex vivo whole human brain at ultra-high gradient strength with structured low-rank echo-planar imaging ghost correction. *NMR Biomed*. 2023;36(2):e4831.

11. Reeder SB, Atalar E, Bolster BD Jr, McVeigh ER. Quantification and reduction of ghosting artifacts in interleaved echo-planar imaging. *Magn Reson Med*. 1997;38(3):429-439.

12. Ahn CB, Cho ZH. A new phase correction method in NMR imaging based on autocorrelation and histogram analysis. *IEEE Trans Med Imaging*. 1987;6(1):32-36.

13. Pan Z, Ma X, Dai E, et al. Reconstruction for 7T high-resolution whole-brain diffusion MRIusing two-stage N/2 ghost correction and L1-SPIRiT without single-band reference. *Magn Reson Med*. 2023;89(5):1915-1930.

14. Yarach U, In MH, Chatnuntawech I, et al. Model-based iterative reconstruction for single-shot EPI at 7T. *Magn Reson Med*. 2017;78(6):2250-2264.

15. Ramos-Llordén G, Park DJ, Kirsch JE, et al. Eddy current-induced artifact correction in

high b-value ex vivo human brain diffusion MRI with dynamic field monitoring. *Magn Reson Med*. 2024;91(2):541-557.

16. Ma R, Akcakaya M, Moeller S, Auerbach E, Ugurbil K, Van de Moortele PF. A field-monitoring-based approach for correcting eddy-current-induced artifacts of up to the 2(nd) spatial order in human-connectome-project-style multiband diffusion MRI experiment at 7T: A pilot study. *Neuroimage*. 2020;216:116861.

17. Buonocore MH, Zhu DC. Image-based ghost correction for interleaved EPI. *Magn Reson Med*. 2001;45(1):96-108.

18. Clare S. Iterative Nyquist ghost correction for single and multi-shot EPI using an entropy measure. In: *Proceedings of the 11th Annual Meeting of ISMRM*. Toronto, ON, Canada; 2003:1041.

19. Chen N, Avram AV, Song AW. Two-dimensional phase cycled reconstruction for inherent correction of echo-planar imaging nyquist artifacts. *Magn Reson Med*. 2011;66(4):1057-1066.

20. Chang HC, Chen NK. Joint correction of Nyquist artifact and minuscule motion-induced aliasing artifact in interleaved diffusion weighted EPI data using a composite two-dimensional phase correction procedure. *Magn Reson Imaging*. 2016;34(7):974-979.

21. Haldar JP. Low-rank modeling of local k-space neighborhoods (LORAKS) for constrained MRI. *IEEE Trans Med Imaging*. 2014;33(3):668-681.

22. Mani M, Jacob M, Yang B, Magnotta V. Comprehensive Correction of Motion and Nyquist Ghost Artifacts for Multi-shot Diffusion Imaging. In: *Proceedings of the 27rd Annual Meeting of ISMRM*. Honolulu, HI, United States; 2017:3498.

23. Lobos RA, Kim TH, Hoge WS, Haldar JP. Navigator-free EPI ghost correction with structured low-rank matrix models: new theory and methods. *IEEE Trans Med Imaging*. 2018;37(11):2390-2402.

24. Hoge WS, Polimeni JR. Dual-polarity GRAPPA for simultaneous reconstruction and ghost correction of echo planar imaging data. *Magn Reson Med*. 2016;76(1):32-44.

25. Griswold MA, Jakob PM, Heidemann RM, et al. Generalized autocalibrating partially parallel acquisitions (GRAPPA). *Magn Reson Med*. 2002;47(6):1202-1210.

26. Hoge WS, Polimeni JR. Artifact Correction in Accelerated-Segmented EPI data via Dual-Polarity GRAPPA. In: Proceedings of the 25th annual meeting of ISMRM. Honolulu, HI, USA; 2017:578.

27. Zhang J, Qian T, Zhang B, Li Q, and Liu W. Multi-Shot Dual-Polarity GRAPPA For Nyquist Ghost And Fuzzy Ripple Artifact Correction Of Echo-Planar Imaging. In: *Proceedings of the 33rd Annual Meeting of ISMRM*. Honolulu, HI, USA; 2025:1364.

28. Layton KJ, Kroboth S, Jia F, et al. Pulseq: A rapid and hardware-independent pulse sequence prototyping framework. *Magn Reson Med*. 2017;77(4):1544-1552.

29. Huang SY, Witzel T, Keil B, et al. Connectome 2.0: Developing the next-generation ultra-high gradient strength human MRI scanner for bridging studies of the micro-, meso- and macro-connectome. *Neuroimage*. 2021;243:118530.

30. Ramos-Llordén G, Lee HH, Davids M, et al. Ultra-high gradient connectomics and microstructure MRI scanner for imaging of human brain circuits across scales. *Nat*

*Biomed Eng*. July 2025. doi:10.1038/s41551-025-01457-x

31. Xiang QS, Ye FQ. Correction for geometric distortion and N/2 ghosting in EPI by phase labeling for additional coordinate encoding (PLACE). *Magn Reson Med*. 2007;57(4):731-741.

32. Pruessmann KP, Weiger M, Scheidegger MB, Boesiger P. SENSE: Sensitivity encoding for fast MRI. *Magn Reson Med*. 1999;42(5):952-962.

33. Chang YV, Zhou K, Hoge WS, Hoelscher U, Liu W, and Polimeni JR. Inter-frame phase alignment for Echo Planar Imaging calibration data acquired with opposite read-out polarities. In: *Proceedings of the 27th Annual Meeting of ISMRM*. Montréal, QC, Canada; 2019:928.

34. Mahmutovic M, Shrestha M, Ramos-Llordén G, Scholz A, Kirsch JE, Wald LL, Möller HE, Mekkaoui C, Huang SY, and Keil B. A 72-Channel Head Coil with an Integrated 16-Channel Field Camera for the Connectome 2.0 Scanner. In: *Proceedings of the 32rd Annual Meeting of ISMRM.* Singapore; 2024:1030.

35. Zhang T, Pauly JM, Vasanawala SS, Lustig M. Coil compression for accelerated imaging with Cartesian sampling. *Magn Reson Med*. 2013;69(2):571-582.

36. Haacke EM, Lindskogj ED, Lin W. A fast, iterative, partial-fourier technique capable of local phase recovery. *J Magn Reson*. 1991;92(1):126-145.

37. Uecker M, Lai P, Murphy MJ, et al. ESPIRiT-an eigenvalue approach to autocalibrating parallel MRI: Where SENSE meets GRAPPA. *Magn Reson Med*. 2014;71(3):990-1001.

38. Kim YC, Nielsen JF, Nayak KS. Automatic correction of echo-planar imaging (EPI) ghosting artifacts in real-time interactive cardiac MRI using sensitivity encoding. *J Magn Reson Imaging*. 2008;27(1):239-245.

39. Wang N, Abraham D, Wu H, Liao C, Cao X, Polimeni J, Huber R, Liu Q, Ning L, Rathi Y, Abad N, Yang B, Kerr A, Westin C, and Setsompop K. Field-correcting GRAPPA (FCG): A generalizable technique to correct spatiotemporal varying odd-even phase errors in EPI, SMS-EPI and 3D-EPI. In: *Proceedings of the 33rd Annual Meeting of ISMRM*. Honolulu, HI, USA; 2025:1363.

40. Bilgic B, Chatnuntawech I, Manhard MK, et al. Highly accelerated multishot echo planar imaging through synergistic machine learning and joint reconstruction. *Magn Reson Med*. 2019;82(4):1343-1358.

41. Liu W, Zhao X, Ma Y, Tang X, Gao JH. DWI using navigated interleaved multishot EPI with realigned GRAPPA reconstruction. *Magn Reson Med*. 2016;75(1):280-286.

42. Ma X, Zhang Z, Dai E, Guo H. Improved multi-shot diffusion imaging using GRAPPA with a compact kernel. *NeuroImage*. 2016;138:88-99.

43. Dai E, Ma X, Zhang Z, Yuan C, Guo H. Simultaneous multislice accelerated interleaved EPI DWI using generalized blipped-CAIPI acquisition and 3D K-space reconstruction. *Magn Reson Med*. 2017;77(4):1593-1605.

44. Lobos RA, Hoge WS, Javed A, et al. Robust autocalibrated structured low-rank EPI ghost correction. *Magn Reson Med*. 2021;85(6):3403-3419.

45. Setsompop K, Gagoski BA, Polimeni JR, Witzel T, Wedeen VJ, Wald LL. Blipped-controlled aliasing in parallel imaging for simultaneous multislice echo planar imaging with reduced g-factor penalty. *Magn Reson Med*. 2012;67(5):1210-1224.

46. Engström M, Skare S. Diffusion-weighted 3D multislab echo planar imaging for high signal-to-noise ratio efficiency and isotropic image resolution. *Magn Reson Med*. 2013;70(6):1507-1514.

47. Poser BA, Koopmans PJ, Witzel T, Wald LL, Barth M. Three dimensional echo-planar imaging at 7 Tesla. *Neuroimage*. 2010;51(1):261-266.

48. Dai E, Liu S, Guo H. High-resolution whole-brain diffusion MRI at 3T using simultaneous multi-slab (SMSlab) acquisition. *Neuroimage*. 2021;237:118099.

49. Hoge WS, Setsompop K, Polimeni JR. Dual-polarity slice-GRAPPA for concurrent ghost correction and slice separation in simultaneous multi-slice EPI. *Magn Reson Med*. 2018;80(4):1364-1375.

50. Hoge WS, Chang Y, Poser BA, Polimeni JR. 3D dual-polarity GRAPPA for ghost correction of volumetric echo-planar imaging data. In: *Proceedings of the 32th Annual Meeting of ISMRM*. Singapore; 2024:4263.

51. Moeller S, Yacoub E, Olman CA, et al. Multiband multislice GE-EPI at 7 tesla, with 16-fold acceleration using partial parallel imaging with application to high spatial and temporal whole-brain fMRI. *Magn Reson Med*. 2010;63(5):1144-1153.

52. Kundu P, Inati SJ, Evans JW, Luh WM, Bandettini PA. Differentiating BOLD and non-BOLD signals in fMRI time series using multi-echo EPI. *Neuroimage*. 2012;60(3):1759-1770.

# Supporting Information

## Dual-polarity SENSE (DPS)

Inspired by the idea of conventional DPG that ghost correction and reconstruction can be integrated into a one-step process, we propose a SENSE-based ghost correction method termed dual-polarity SENSE (DPS). The central concept of DPS is to combine ghost correction with SENSE reconstruction. To achieve that, the phase difference between RO+ and RO- data is embedded within a "dual-polarity" sensitivity map, which is subsequently used for ghost-free image reconstruction.

Figure S2 illustrates the DPS workflow. For calibration, DPS uses the same calibration data as DPG, which comprises a pair of fully sampled RO+ and RO- frames. These two frames are concatenated along the channel dimension, and ESPIRiT is employed to calculate the dual-polarity sensitivity map. For reconstruction, the RO+ and RO- parts of the imaging data are separated and similarly concatenated in the channel dimension, forming undersampled dual-polarity data. The undersampled data is then reconstructed by SENSE, using the previously calculated dual-polarity sensitivity map. Similar to msDPG, DPS also facilitates regularized reconstruction, as it is SENSE-based and inherently compatible with advanced regularizers.

### (a) Dual-polarity SENSE calibration

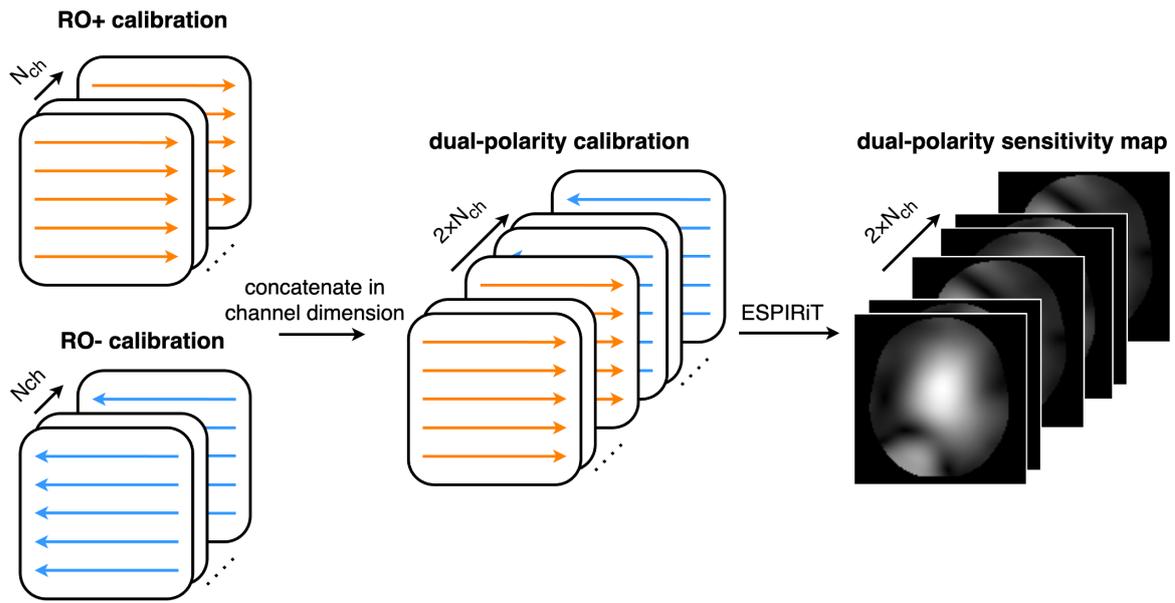

### (b) Dual-polarity SENSE reconstruction

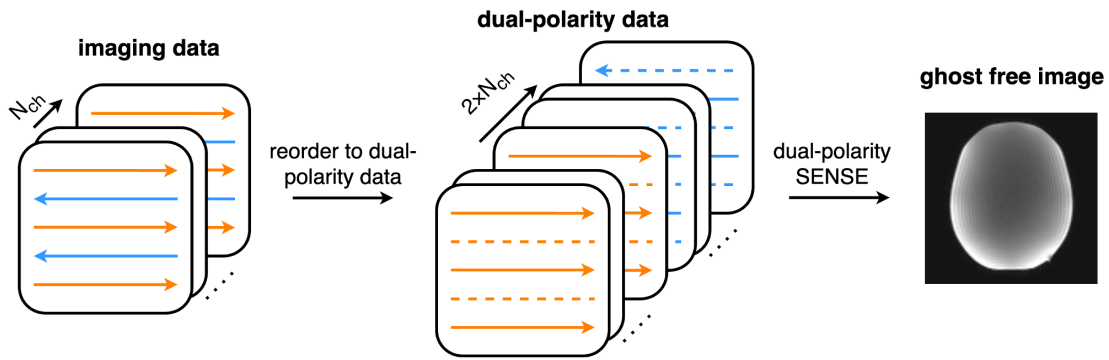

**Figure S1** Illustration of dual-polarity SENSE (DPS) calibration and reconstruction. In each k-space frame, solid and dashed lines represent acquired and unacquired phase encoding lines, respectively. Arrow direction in the solid lines indicates the readout direction: rightward arrows represent positive readout (RO+), and leftward arrows indicate negative readout (RO-). (a) DPS calibration. DPS uses the same calibration data as dual-polarity GRAPPA (DPG), comprising both RO+ and RO- data, each with $N_{ch}$ channels. The RO+ and RO- data are concatenated along the channel dimension, forming dual-polarity calibration data with $2 \times N_{ch}$ channels. ESPIRiT is applied to calculate the dual-polarity sensitivity map. (b) DPS reconstruction. The acquired EPI imaging data are first separated and then concatenated to generate undersampled dual-polarity data, which are SENSE reconstructed by the previously calculated dual-polarity sensitivity map.

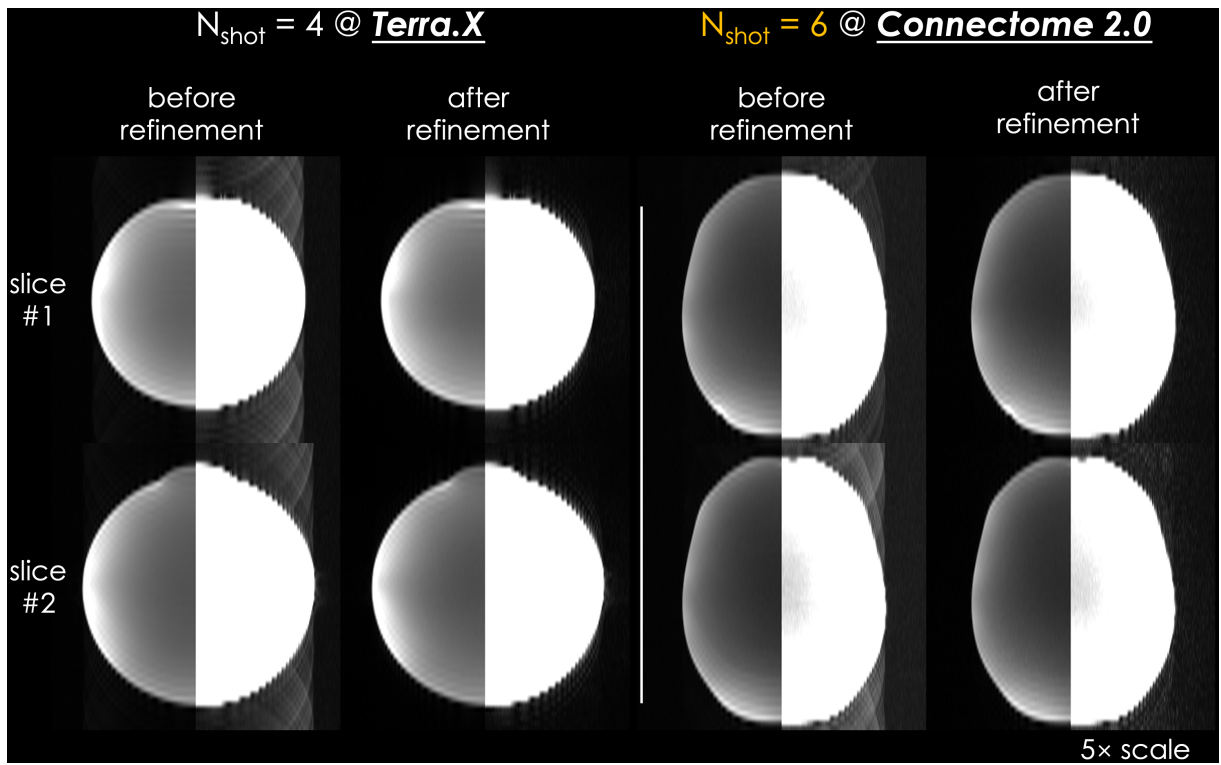

**Figure S2** Positive readout (RO+) phantom calibration images before and after the calibration refinement. The left panel shows images acquired on the Terra.X scanner using a 4-shot EPI sequence, and the right panel shows images from the Connectome 2.0 scanner using a 6-shot acquisition. Two representative slices are shown in each panel, and the right half of each slice is intensified fivefold to emphasize ghosts. In each panel, the images before calibration refinement contain noticeable residual artifacts, while those after refinement are ghost-free. In the $N_{shot}$ = 6 case (the right panel), the images after calibration refinement maintain a satisfying signal-to-noise ratio (SNR), thanks to the "shot combination" strategy.

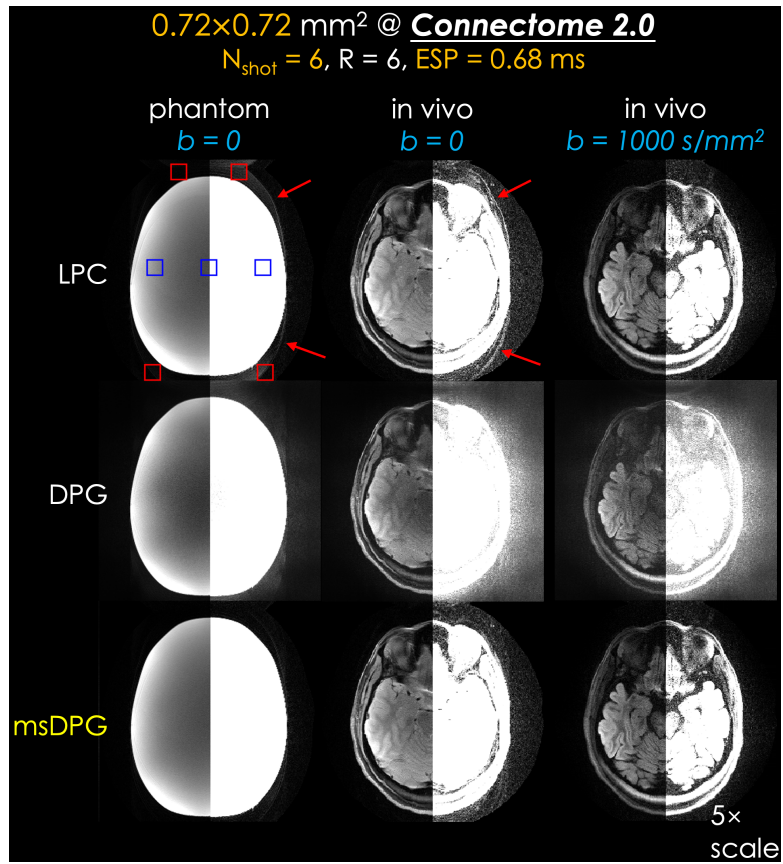

**Figure S3** Submillimeter phantom and in vivo images reconstructed by different methods. Data were acquired from the Connectome 2.0 scanner using a 6-shot multishot EPI sequence. Phantom (b = 0 s/mm$^2$), in vivo (b = 0 s/mm$^2$), and in vivo (b = 1000 s/mm$^2$) results are displayed from left to right. The right half of each image is intensified fivefold to emphasize ghosts. The blue and red rectangles in the phantom images represent ROIs for signal and ghost, respectively. Conventional DPG exhibits compromised performance in this case, producing noise-contaminated images as it reconstructs each shot separately with a high in-plane acceleration factor (R = 6). msDPG offers superior SNR than DPG, effectively suppressing the ghosts observed in the LPC images (arrows).

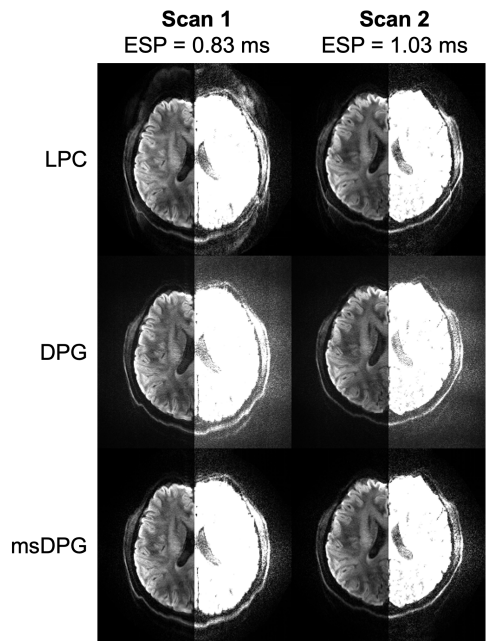
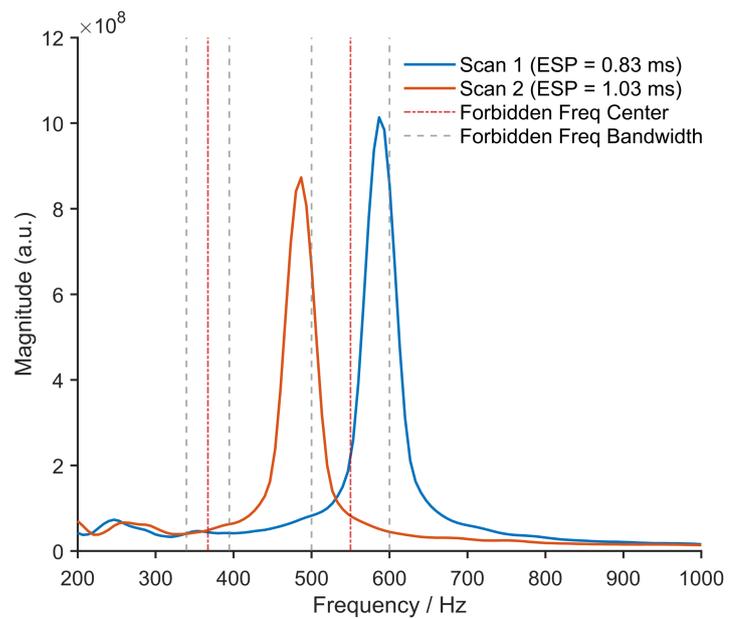

**Figure S4** Comparison of two scans with different echo spacings (ESPs). Both scans were conducted on the Terra.X scanner using a 3-shot EPI sequence. Scan 1 has an ESP of 0.83 ms, while Scan 2 has an ESP of 1.03 ms. (a) Image reconstruction results. The LPC results of Scan 1 exhibit severe ghosts, while those of Scan 2 contain fewer ghosts. In both cases, msDPG and DPG effectively mitigate residual ghosts, while DPG images show higher noise levels. (b) Calculated gradient resonance spectrum of the two scans. The blue and red lines represent the spectra of Scans 1 and 2, respectively. The dashed lines indicate the forbidden frequency range. The spectrum of Scan 1 has a peak in the forbidden frequency bandwidth.

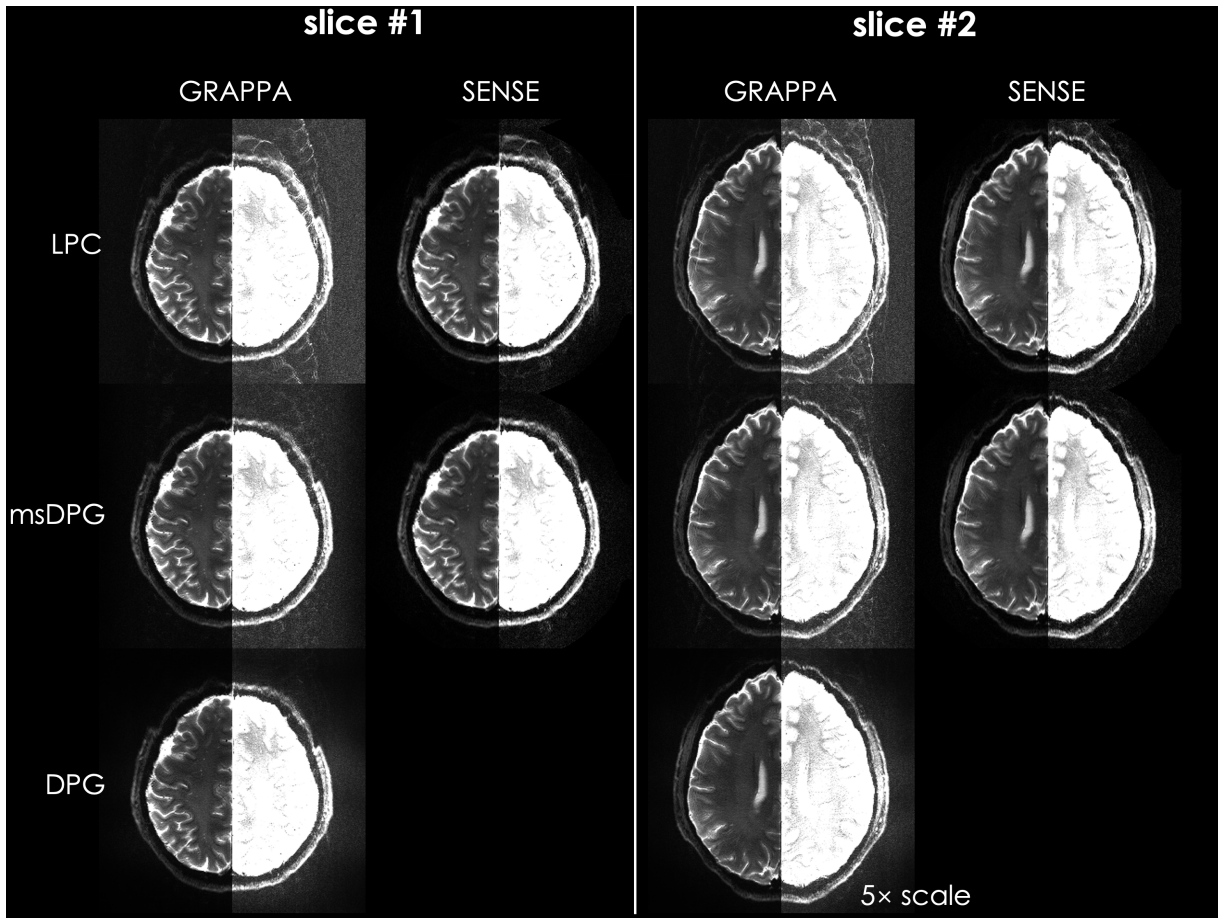

**Figure S5** Comparison of GRAPPA and SENSE for b = 0 s/mm$^2$ image reconstruction. Two slices from the in vivo b = 0 s/mm$^2$ dataset acquired using 4-shot EPI on Terra.X are presented in the left and right panels. The right half of each image is intensified fivefold to emphasize ghosts. The images are reconstructed by LPC, msDPG, and DPG via GRAPPA or SENSE-based methods. DPG is incompatible with SENSE because it is intrinsically GRAPPA-based. LPC images contain noticeable artifacts, while the images of msDPG and DPG are mainly ghost-free. The results of SENSE exhibit fewer artifacts than those of GRAPPA.

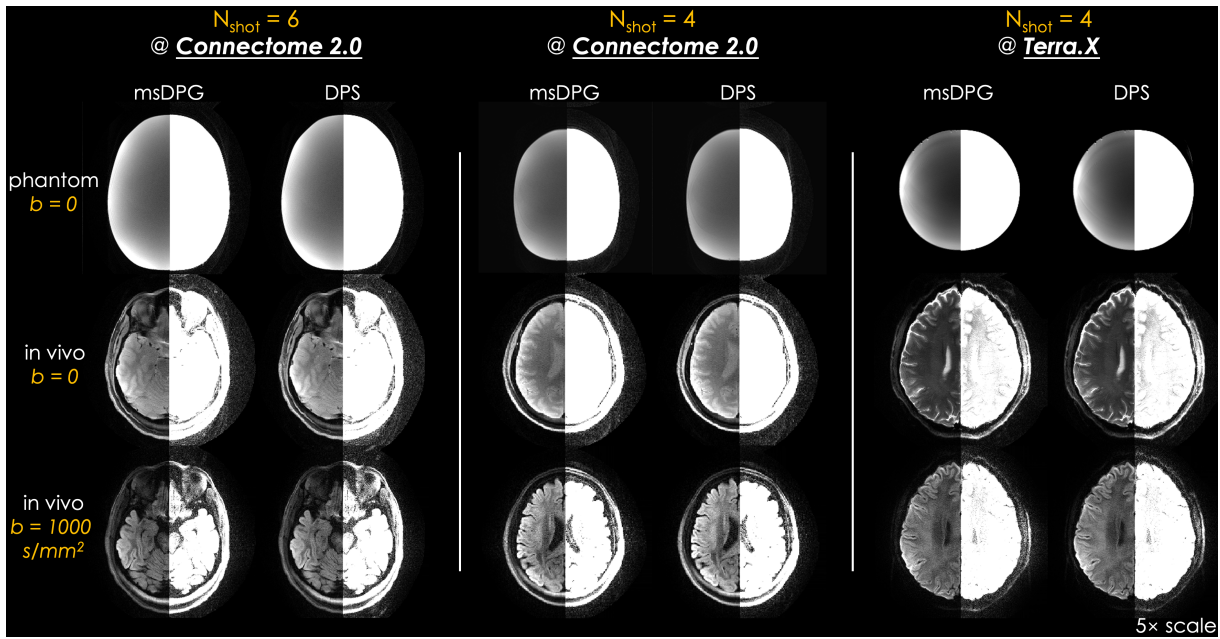

**Figure S6** Comparison of msDPG and DPS reconstruction results on different datasets. The left and middle panels show results from the 6-shot and 4-shot scans at Connectome 2.0, and the right panel shows results from the 4-shot scan at Terra.X. In each panel, phantom (b = 0 s/mm$^2$), in vivo (b = 0 s/mm$^2$), and in vivo (b = 1000 s/mm$^2$) results are respectively displayed from top to bottom. The right half of each image is intensified fivefold to emphasize ghosts. DPS exhibits a similar ghost correction performance to msDPG.